\documentclass[]{aipproc}

\layoutstyle{6x9}

\SetInternalRegister\hbadness{8000} % pseudo latin isn't breaking very well :-)  

\begin{document}

\title 
      [Neutrino Mass Bounds]
      {Neutrino Mass Bounds 
from $0\nu\beta\beta$ Decays and Large Scale Structures}

\classification{98.80.Jk, 98.80.Cq, 98.80.-k}
\keywords{Neutrino Masses, Interacting Neutrino Dark-Energy, Neutrinoless Double Beta Decays}

\author{Y.-Y. Keum}{
  address={Theory Division, National Astronomical Observatory, Mitaka, Tokyo 181-8588, Japan},
  email={yykeum@phys.ntu.edu.tw},
altaddress={Department of Physics, National Taiwan University, Taipei, Taiwan 10672, R.O.C.},
}\footnote{Talk presented by Y.-Y. Keum at the 10th international Symposium  on
Origin of Matter and Evolution of Galaxies, Sapporo, Japan, Dec. 4-7 2007.\\
Email: yykeum@phys.ntu.edu.tw}

\iftrue
\author{K. Ichiki}{
  address={Research Center for the Early Universe, University of Tokyo, 7-3-1
Hongo, Bunkyo-ku, Tokyo 113-0033, Japan},
  email={ichiki@resceu.s.u-tokyo.ac.jp},
}

\author{T. Kajino}{
  address={Theory Division, National Astronomical Observatory, Mitaka, Tokyo 181-8588, Japan},
  email={kajino@nao.ac.jp},
 % homepage={http://www.dcarlisle.demon.co.uk},
 altaddress={Department of Astronomy, Graduate School of Science, University of Tokyo, Bunkyo-ku, Tokyo 113-0033, Japn}
}
\fi

% \copyrightholder{Acoustical Scociety of America}
\copyrightyear  {2008}

\begin{abstract}
We investigate the way how the total mass sum of neutrinos
can be constrained from the neutrinoless double beta decay and cosmological probes
with cosmic microwave background (WMAP 3-year results), large scale structures including 2dFGRS and SDSS data sets.
First we discuss, in brief, on the current status of neutrino mass bounds from 
neutrino beta decays and cosmic constrain within the flat $\Lambda CMD$ model.  
In addition, we explore the interacting neutrino dark-energy model, where the evolution of neutrino masses  is determined by quintessence scalar filed, which is responsable for cosmic acceleration today.                         
Assuming the flatness
of the universe, the constraint we can derive from the current
observation is $\sum m_{\nu} < 0.87$eV at the 95 $\%$ confidence
level, which is consistent with  $\sum m_{\nu} < 0.68$eV in the flat $\Lambda CDM$
model. Finally we discuss the future prospect of the neutrino mass bound with weak-lensing effects.
\end{abstract}

\date{\today}

\maketitle

\section{Neutrino Mass Bounds from the $0\nu\beta\beta$ Decays and Cosmological Probes with $\Lambda CMD$ model}
The existence of the tiny neutrino masses qualifies as the first evidence of
new physics beyond the Standard Model. The answers to the hot questions on
(1) whether neutrinos are Dirac or Majorana fermions ?,
(2) the mass hierarchy pattern (normal or inverted hierarchy type ?),
(3) the absolute value of the neutrino mass, 
will provide us the additional knowledge about the precise nature of this new physics, have the potential to unravel some of the deepest and most long-standing mysteries of cosmology and astrophysics, such as the origin of matter, 
the origin of heavy elements, and even the nature of dark-energy.

The golbal analysis of the solar and KamLAND data \cite{solar-data}
and super-Kamiokande atmospheric data \cite{super-kam}
provide the two independent neutrino mass-squared differences:
$\Delta m_{sol}^2 \equiv \Delta m_s^2 = (7.9^{+2.8}_{-2,9}) \times 10^{-5} eV^2$,
$|\Delta m_{atm}^2| \equiv |\Delta m_a^2| = (2.6 \pm 0.2) \times 10^{-3} eV^2$,
and mixing angles: $\theta_{12}=\theta_3 =(33.7 \pm 1.3)^o, 
\theta_{23}=\theta_1=(43.3^{+4.3}_{-3.8})^o$ and $\theta_{13}=\theta_2 < 5.2^o$. 
Above results tell us a substential evidence that the three known neutrinos have a combined mass ($\Sigma =\sum_{i=1}^{i=3} m_{\nu,i}$) of at least $\sqrt{|\Delta m_a|^2} \sim 0.05 \, eV$. 
Since neutrino oscillations are only sensitive to mass-squared differences, 
three possible neutrino mass spectrum are allowed as:
$m_1 \ll m_2 \ll m_3$ (normal hierarchy), or
$m_3 \ll m_1 \ll m_2$ (inverted hierarchy), or
$m_1 \simeq m_2 \simeq m_3$ (quasi-degenerate),
depending on whether the light eigenmass is close to 0 or 
$\gg \sqrt{|\Delta m_a^2|}$, respectively.

The nature of spectrum is important to neutrino mass model-building,
the combination of neutrinos to dark-matter, and the viability of observing neutrinoless double beta-decay ($0\nu\beta\beta$) if neutrinos are Majorana
\cite{{zeldovich}, {barger:2002}, {bilenky:2004}}. 

There are three well
known ways to get the direct information on the absolute mass of
neutrinos by using: Tritium $\beta$-decay experiment, neutrinoless
double beta decay experiment, and astrophysical observations.

\subsection{(A) Neutrinoless Double Beta Decays:}
The standard method for the measurement of the absolute value of the
neutrino mass is based on the detailed investigation of the
high-energy part of the $\beta$-spectrum of the decay of tritium:
\begin{equation}
{}^3H \longrightarrow {}^3He + e^{-} + \bar{\nu}_e
\end{equation}
This decay has a small energy release ($E_0 \simeq 18.6 keV$) and a
convenient life time ($T_{1/2} = 12.3 $ years). Since the flavour
eigenstates are different from mass eigenstates in neutrino sector,
in general, electron neutrino can be expressed as
\begin{equation}
\nu_{eL} = \sum_{i} U_{ei} \, \nu_{iL},
\end{equation}
where $\nu_i$ is the field of neutrino with mass $m_i$, and U is the
unitary mixing matrix. Neglecting the recoil of the final nucleus,
the spectrum of the electrons is given:
\begin{equation}
{d\Gamma \over d E} = \sum_{i} |U_{ei}|^2 \,\, {d\Gamma_{i} \over
dE},
\end{equation}
and the resulting spectrum can be analyzed in term of a single
mean-squared electron neutrino mass
\begin{equation}
\langle m_{\beta} \rangle^2 =\sum_{j} m_{j}^2|U_{ej}|^2 =
m_1^2|U_{e1}|^2 + m_2^2|U_{e2}|^2 + m_3^2 |U_{e3}|^2
\end{equation}
If the neutrino mass spectrum is practically degenerate: $m_1 \simeq
m_2 \simeq m_3$, the neutrino mass can be measured in these
experiments. Present-day tritium experiments Mainz\cite{mainz} and
Troitsk\cite{troitsk} gave the following results:
\begin{eqnarray}
m_1^2 &=& (-1.2\pm 2.2 \pm 2.1) \, eV^2 \hspace{5mm} {\rm (Mainz)},
\\
&=& ( -2.3 \pm 2.5 \pm 2.0) \, eV^2 \hspace{5mm}  {\rm (Troitsk)}.
\end{eqnarray}
This value corresponds to the upper bound
\begin{equation}
m_1 < \, 2.2 eV  \hspace{3mm} (95 \% C.L.)
\end{equation}
Another useful method is by using the neutrinoless double beta
decay. The search for neutrinoless double $\beta$-decay
\begin{equation}
(A,Z) \longrightarrow (A,Z_2) + e^- + e^-
\end{equation}
for some even-even nuclei is the most sensitive and direct way of
investigating the nature of neutrinos with definite masses. In this
process, total lepton number is violated ($\Delta L =2$) and is allowed 
only if massive neutrinos are Majorana particles. The rate of
$0\nu\beta\beta$ is approximately
\begin{equation}
{1 \over T^{0\nu}_{1/2}} = G_{0\nu}(Q_{\beta\beta}, Z) \,
|M_{0\nu}|^2 \,\, \langle m_{\beta\beta} \rangle^2,
\end{equation}
where $G^{0\nu}$ is the phase space factor for the emission of the
two electrons, $M_{0\nu}$ is nuclear matrix elements, and
$<m_{\beta\beta}> $ is the effective majorana mass of the electron
neutrino:
\begin{equation}
\langle m_{\beta\beta} \rangle \equiv |\sum_{i} U_{ei}^2 m_i|
\label{eq:mbetabeta}
\end{equation}
We can write eq.(\ref{eq:mbetabeta}), for normal and inverted
hierarchy respectively, in terms of mixing angles and
$\Delta_s^2=m_2^2 -m_1^2 = (7.9^{+2.8}_{-2.9}) \cdot 10^{-5} \, eV^2$, 
$\Delta_a= \pm (m_3^2 -m_2^2) \simeq \pm(2.6 \pm 0.2) \cdot 10^{-3} \, eV^2$ 
and CP phases as follows:
\begin{eqnarray}
\langle m_{ee} \rangle &=& \left|c_2^2 c_3^2 m_1 + c_2^2 s_3^2
e^{i\phi_2} \sqrt{\Delta_s^2 + m_1^2} + s_2^2 e^{i\phi_3}
\sqrt{\Delta_a^2 +
m_1^2} \right|, \hspace{3mm} {\rm (normal \,\, hyerarchy)}; \nonumber \\
\langle m_{ee} \rangle &=& \left|s_2^2 m_1 + c_2^2 s_3^2 e^{i\phi_2}
\sqrt{\Delta_a^2 - \Delta_s^2 + m_1^2} + c_2^2s_2^2 e^{i\phi_3}
\right|, \hspace{6mm} {\rm (inverted \,\, hyerarchy)}.
\end{eqnarray} 
From above relations, we can have the correlation plot between $m_{light}$ 
and $|m_{\beta\beta}|$ with current observed data sets of mixing angles and
$\Delta_{s,a}^2$ from neutrino oscillation experiments.
However, $0\nu\beta\beta$ decays have not yet been seen experimentally. 

%%%%%%%%%%%%%%%%%%%%%%%%%%%%%%%%%%%%%%%%%%%%
The most stringent lower bounds for the time of life of
$0\nu\beta\beta$-decay were obtained in the
Heidelberg-Moscow\cite{heidelberg-moscow} and IGEX\cite{igex}
${}^{76} Ge$ experiments:
\begin{eqnarray}
&& T^{0\nu}_{1/2} \geq 1.9 \cdot 10^{25} years \hspace{5mm} (90 \%
C.L.) \hspace{5mm} {\rm Heidelberg-Moscow}, \\
&& T^{0\nu}_{1/2} \geq 1.57 \cdot 10^{25} years \hspace{5mm} (90 \%
C.L.) \hspace{5mm} {\rm IGEX}.
\end{eqnarray}
%%%%%%%%%%%%%%%%%%%%%%%%%%%%%%%%%%%%%%%%%%%%%%%%%%%%%%%%%%%%%%
\begin{table}
\begin{tabular}{cc|ccc}
\hline 
Nucl.~~ & ~~$M^{0\nu}$~~ & ~~$T^{0\nu}_{1/2}$ (years)~~ & ~~Experiment~~ & ~~$|m_{\beta\beta}|$ (eV)~~ \\
\hline 
${}^{76}Ge$& 2.40  & $1.9 \, \cdot \, 10^{25}$ & Hiedelberg-Moscow & 0.55 \\
           &       & $3 \, \cdot \, 10^{27}$ & Majorana & 0.044 \\ 
           &       & $7 \, \cdot \, 10^{27}$ & GEM & 0.028 \\  
           &       & $1 \, \cdot \, 10^{28}$ & GENIUS & 0.023 \\
\hline
${}^{100}Mo$& 1.16 & $6.0 \, \cdot \, 10^{22}$ & NEM03 & 7.8 \\
          &        & $4 \, \cdot \, 10^{24}$ & NEM03 & 0.92 \\ 
          &        & $1 \, \cdot \, 10^{27}$ & MOON & 0.058 \\  
\hline
${}^{130}Te$ & 1.50 & $1.4 \, \cdot \, 10^{23}$ & CUORE & 3.9 \\
            &       & $2 \, \cdot \, 10^{26}$ & CUORE & 0.10 \\         
\hline
${}^{136}Xe$ & 0.98 & $1.2 \, \cdot \, 10^{24}$ & DAMA & 2.3 \\
            &      & $3 \, \cdot \, 10^{26}$ & XMASS & 0.10 \\ 
            &      & $2 \, \cdot \, 10^{27}$ & EXO(1t) & 0.055 \\  
            &      & $4 \, \cdot \, 10^{28}$ & EXO(10t) & 0.012 \\
\hline 
\end{tabular}
%\source{Central Statistical Office, UK}
\caption{The current upper limits on effective Majorana neutrino mass $|m_{\beta\beta}|$ and the sensitivities of the future $0\nu\beta\beta$-decay experiments.
We used the matrix elements $M^{0\nu}$ with reduced uncertainty \cite{Rodin:2003}.
$T^{0\nu}_{1/2}$ denotes the current lower limit on the $0\nu\beta\beta$-decay half-life or the sensitivity of planned $0\nu\beta\beta$-decay experiments.
}
\label{tab:upperlimit}
\end{table}
%%%%%%%%%%%%%%%%%%%%%%%%%%%%%%%%%%%%%%%%%%%%%%%%%%%%%%%%%%%%%%%%%%%%
Taking into account different calculation of the nuclear matrix
elements, from these results the following upper bounds were
obtained for the effective Majorana mass:
\begin{equation}
|m_{\beta\beta}| \, < \, (0.35 - 1.24) \, eV
\end{equation}
Many new experiments (including CAMEO, CUORE, COBRA, EXO, GENIUS,
MAJORANA, MOON and XMASS experiments) on the search for the
neutrinoless double $\beta$-decay are in preparation at present. In
these experiments the sensitivities
\begin{equation}
|m_{\beta\beta}| \simeq (0.1 - 0.015)\, eV
\end{equation}
are expected to be achieved. 
The detail upper limit of $|m_{\beta\beta}|$ and the sensitivities of the
future $0\nu\beta\beta$-decay experiments are summerized 
in table \ref{tab:upperlimit}. It is very difficult to confirm
the normal hierarchy pattern of neutrino mass when $m_1 < 1.7 \cdot 10^{-3} \, eV $, however for the inverted case, it can be detected if  $m_3 < 8.9 \cdot 10^{-3} \, eV$
and $m_{ee} > 0.012 \, eV$.
%%%%%%%%%%%%%%%%%%%%%%%%%%%%%%%%%%%%%%%%%%%%%%%%%%%%%%%%%%%%%%
%%%%%%%%%%%%%%%%%%%%%%%%%%%%%%%%%%%%%%%%%%%%%%%%%%%%%%%%%%%%%%%%%
\subsection{(B) Cosmological Constrains within the Standard Cosmology:}
Within the standard cosmological model, the relic abundance of
neutrinos at present epoch was come out straightforwardly from the
fact that they follow the Fermi-Dirac distribution after freeze
out, and their temperature is related to the CMB radiation
temperature $T_{CMB}$ today by $T_{\nu} =(4/11)^{1/3} T_{CMB}$ with
$T_{CMB}=2.726$ K, providing
\begin{equation}
n_{\nu} = { 6 \zeta(3) \over 11 \pi^2} \, T_{CMB}^3,
\label{eq:nu-density}
\end{equation}
where $\zeta(3)\simeq 1.202$, which gives $n_{\nu} \simeq 112
cm^{-3}$ for each family of neutrinos at present. By now the massive
neutrinos become non-relativistic, and their contribution to the
mass density ($\Omega_{\nu}$) of the universe can be expressed as
\begin{equation}
\Omega_{\nu} h^2 ={\Sigma \over 93.14 eV}. \label{eq:omega-nu}
\end{equation}
where $\Sigma$ stands for the sum of the neutrino masses.
In this relation, we include the effect of three neutrino oscillation
\cite{mangano:2005}.
We should notice that when obtaining the limit of neutrino masses
one usually assumes:
\begin{itemize}
\item the standard spatially flat $\Lambda CDM$ model with adiabatic
primordial perturbations,
\item they have no non-standard interactions,
\item neutrinos decoupled from the thermal background at the
temperatures of order 1 MeV.
\end{itemize}
These simple conditions can be modified from several effects: due
to a sizable neutrino-antineutrino asymmetry, due to additional
light scalar field coupled with neutrinos \cite{beacom:2004}, and due
to the light sterile neutrino \cite{dodelson:2006}. However, analysis
of WMAP and 2dFGRS data gave independent evidence for small lepton
asymmetries \cite{{hannestad:2003},{pierpaoli:2003}}, and  such a
scenario with a light scalar field  is strongly disfavored by the
current CMB power spectrum data \cite{hannestad:2005}. We will not
therefore take into account such non-standard couplings of neutrinos
in the following. In addition, current cosmological observations are
sensitive to neutrino masses $0.1 \,{\rm eV} \, < \, \Sigma \, <\,
2.0 \, {\rm eV}$. In this mass scale, the mass-square differences
are small enough and all three active neutrinos are nearly
degenerate in mass. Therefore we take the assumption of degenerate
mass hierarchy. Even if we consider different mass hierarchy
pattern, it will be very difficult to distinguish such hierarchy
patterns from cosmological data alone
%{\it since the total mass of
%neutrino is only sensitive to the large scale structure and CMB data
%at present} 
\cite{slosar:2006}.

After neutrinos decoupled from the thermal background, they stream freely
and their density perturbations are damped on scale smaller than their
free streaming scale. Consequently the perturbations of cold 
dark matter (CDM) and baryons grow more slowly because of the missing
gravitational contribution from neutrinos. The free streaming scale
of relativistic neutrinos grows with the hubble horizon. When the
neutrinos become non-relativistic, their freestreaming scale shrinks,
and they fall back into the potential wells. The neutrino density
perturbation with scales larger than the freestreaming scale resumes to
trace those of the other species.  
Thus the free streaming effect suppresses the power spectrum on scales
smaller than the horizon when the neutrinos become
non-relativistic. 
The co-moving wavenumber corresponding to this scale is given by
\begin{equation}
k_{nr} = 0.026 \,\left( {m_{\nu} \over 1 \, eV} \right)^{1/2} \,
\Omega_m^{1/2} \, h \, {\rm Mpc}^{-1}, \label{jeans-length}
\end{equation}
for degenerated neutrinos, with almost same mass $m_{\nu}$. The
growth of fourier modes with $k > k_{nr}$ will be suppressed because
of neutrino free-streaming. The power spectrum of matter
fluctuations can be written as
\begin{equation}
P_m(k,z) = P_{*}(k) \, T^2(k,z), \label{eq:power-spectrum}
\end{equation}
where $P_{*}(k)$ is the primordial spectrum of matter fluctuations,
to be a simple power law $P_{*}(k) = A \, k^n$, where A is the
amplitude and n is the spectral index. 
%Here the transfer functions
%of the perturbation T(k,z) is inversely proportional to linear
%growth factor D(z). 
Here the transfer function $T(k,z)$ represents the evolution of
perturbation relative to the largest scale. If some fraction of the
matter density (e.g., neutrinos or dark energy) is unable to cluster,
the speed of growth of perturbation 
becomes slower. 
Because the contribution to the fraction of matter density from
neutrinos is propotional to their masses (Eq. (\ref{eq:omega-nu})), 
the larger mass leads to the smaller growth of perturbation.
%Therefore when neutrinos become more massive,
%the linear growth factor increases at the smaller scale (larger k),
%and simultaneously the power spectrum of the matter fluctuations is
%suppressed. 
The suppression of the power spectrum on small scales is
roughly proportional to $f_{\nu}$ \cite{Hu:1997mj}:
\begin{equation}
{ \Delta P_m(k) \over P_m(k)} \simeq -8 f_{\nu}.
\label{eq:power-spectrum-02}
\end{equation}
where $f_{\nu}=\Omega_{\nu}/\Omega_{M}$ is the fractional
contribution of neutrinos to the total matter density. This result
can be understood qualitatively from the fact that only a fraction
$(1-f_{\nu})$ of the matter can cluster when massive neutrinos are
present \cite{silk:1980}.
%%%%%%%%%%%%%%% Table of EoS %%%%%%%%%%%%%%%%%%%%%%%%%%%%%%%%%%%%
\begin{table}
\caption{Recent cosmological neutrino mass bounds ($95 \%$ C.L.)}
{\begin{tabular}{c||c||c} \hline
Cosmological Data Set & $\Sigma$ bound ($2 \sigma$) & References \\
\hline 
CMB (WMAP-3 year alone) & $< \, 2.0$ eV & Fukugita et al.\cite{fukujita:2006} \\
LSS[2dFGRS] & $ < \, 1.8 $ eV & Elgaroy et al.\cite{elgaroy:2002} \\
CMB + LSS[2dFGRS] & $< \, 1.2$ eV & Sanchez et al.\cite{sanchez:2005} \\
              "        &  $< \, 1.0 $ eV  &
              Hannestad\cite{hannestad:2004} \\
CMB + LSS + SN1a & $< \, 0.75 $ eV & Barger et al.\cite{barger:2003} \\
       "            & $< \, 0.68 $ eV & Spergel et al.\cite{spergel:2006} \\
CMB + LSS + SN1a + BAO & $< \, 0.62 $ eV & Goobar et al.\cite{goobar:2006} \\
        "               & $< \, 0.58 $ eV &                              \\
CMB + LSS + SN1a + Ly-$\alpha$ & $< \, 0.21 $ eV &    Seljak et al.\cite{seljak:2006} \\
CMB + LSS + SN1a + BAO + Ly-$\alpha$ & $ < \, 0.17 $ eV & Seljak et
al.\cite{seljak:2006} \\ \hline
\end{tabular} \label{table:neutrino-mass-bound}}
\end{table}
%%%%%%%%%%%%%%%%%%%%%%%%%%%%%%%%%%%%%%%%%%%%%%%%%%%%%%%%%%%%%%%
Analyses of CMB data are not sensitive to neutrino masses 
if neutrinos behave as massless particles at the epoch of last scattering.
%due to the
%fact that eV mass neutrino behave essentially like cold dark matter
%at the epoch of last scattering. 
According to the analytic
consideration in \cite{ichikawa:2005}, since the redshift when
neutrino becomes non-relativistic is given by $1+z_{nr}=6.24 \cdot
10^4 \, \Omega_{\nu} \, h^2$ and $z_{rec}=1088$, neutrinos become
non-relativistic before the last scattering 
when $\Omega_{nu}h^2 > 0.017$ (i.e. $\Sigma > 1.6 e V$). 
Therefore the dependence of the position of the first peak
and the height of the first peak on $\Omega_{\nu}h^2$ has a
turning point at $\Omega_{\nu}h^2 \simeq 0.017$. This value also
affects CMB anisotropy via the modification of the integrated
Sachs-Wolfe effect due to the massive neutrinos.
 However an important role of CMB data
is to constrain  other parameters that are degenerate with $\Sigma$.
Also, since there is a range of scales common to the CMB and LSS
experiments, CMB data provides an important constraint on the bias
parameters. We summarize some of the recent cosmological neutrino
mass bounds within the flat-$\Lambda CDM$ model 
in table \ref{table:neutrino-mass-bound}.
%%%%%%%%%%%%%%%%%%%%%%%%%%%%%%%%%%%%%%%%%%%%%%%%%%%%%%%%%%%%%%%%%%%

\section{Neutrino Mass Bounds in Interacting Neutrino-Dark Energy Model}
With our previous works
\cite{{ichiki-yyk:2007a},{yyk-2007},{ichiki-yyk:2008a}}, 
we investigate 
the cosmological
implication of an idea of the dark-energy interacting with neutrinos
\cite{{mavanu},{Fardon:2003eh}}. For simplicity, we consider the
case that dark-energy and neutrinos are coupled such that the mass
of the neutrinos is a function of the scalar field which drives the
late time accelerated expansion of the universe.

In our scenario, 
Equations for quintessence scalar field are given by
\begin{eqnarray}
\ddot{\phi}&+&2{\cal H}\dot\phi+a^2\frac{d V_{\rm eff}(\phi)}{d\phi}=0~,
 \label{eq:Qddot}\\
V_{\rm eff}(\phi)&=&V(\phi)+V_{\rm I}(\phi)~,\\
V_{\rm I}(\phi)&=&a^{-4}\int\frac{d^3q}{(2\pi)^3}\sqrt{q^2+a^2
 m_\nu^2(\phi)}f(q)~,\\
m_\nu(\phi) &=& \bar m_i e^{\beta\frac{\phi}{M_{\rm pl}}}~,
\end{eqnarray}
where $V(\phi)$ is the potential of quintessence scalar field, $V_{\rm
I}(\phi)$ is additional potential due to the coupling to neutrino
particles \cite{Fardon:2003eh,Bi:2003yr},
and $m_\nu(\phi)$ is the mass of neutrino coupled to the scalar field,
where we assume the exponential coupling with a coupling parameter
$\beta$. 
${\cal H}$ is $\frac{\dot a}{a}$, where the dot represents the
derivative with respect to the conformal time $\tau$.

Energy densities of mass varying neutrino (MaVaNs) and quintessence scalar
field are described as
\begin{eqnarray}
\rho_\nu &=& a^{-4}\int \frac{d^3 q}{(2\pi)^3} \sqrt{q^2+a^2m_\nu^2} f_0(q)~, \label{eq:rho_nu}\\
3P_\nu &=& a^{-4}\int \frac{d^3 q}{(2\pi)^3} \frac{q^2}{\sqrt{q^2+a^2m_\nu^2}}
 f_0(q)~, \label{eq:P_nu}\\
\rho_\phi &=& \frac{1}{2a^2}\dot\phi^2+V(\phi)~,\label{eq:rho_phi}\\
P_\phi &=& \frac{1}{2a^2}\dot\phi^2-V(\phi)~.
\end{eqnarray}
From equations (\ref{eq:rho_nu}) and (\ref{eq:P_nu}), the equation of
motion for the background energy density of neutrinos is given by
\begin{equation}
\dot\rho_{\nu}+3{\cal H}(\rho_\nu+P_\nu)=\frac{\partial \ln
 m_\nu}{\partial \phi}\dot\phi(\rho_\nu -3P_\nu)~.
\end{equation}
Here we consider three different types of the quintessence potential $V(\phi)$:
(1) inverse power law potentials (Model I), 
(2) SUGRA type potential models (Model II),
(3) exponential type potentials (Model III),
which are given by, respectively:
\begin{equation}
V(\phi)=M^{4}\left({M_{pl}\over \phi}\right)^{\alpha}~ \hspace{3mm}; \hspace{3mm}
M^{4} \left({M_{pl}\over \phi}\right)^{\alpha} e^{3\phi^2/2M_{\rm
pl}^2}~\hspace{3mm}; \hspace{3mm}
M^4 e^{-\alpha ({\phi \over M_{pl}})}.
\end{equation}
The coupling between
cosmological neutrinos and dark energy quintessence could modify the
CMB and matter power spectra significantly.  It is therefore
possible and also important to put constraints on coupling
parameters from current observations. For this purpose, we use the
WMAP3 \cite{Hinshaw:2006ia,Page:2006hz} and 2dFGRS \cite{Cole:2005sx}
data sets.

The flux power spectrum of the Lyman-$\alpha$ forest can be used to
measure the matter power spectrum at small scales around $z < 3$ \cite{McDonald:1999dt,Croft:2000hs}.
It has been shown, however, that the resultant constraint on neutrino
mass can vary significantly from $\sum m_\nu < 0.2$eV to $0.4$eV
depending on the specific Lyman-$\alpha$ analysis used \cite{Goobar:2006xz}.
The complication arises because the result suffers from the
systematic uncertainty regarding to the model for the intergalactic
physical effects, i.e., damping wings, ionizing radiation fluctuations,
galactic winds, and so on \cite{McDonald:2004xp}.
Therefore, we conservatively omit the Lyman-$\alpha$ forest data from
our current analysis.

Because there are many other cosmological parameters than the MaVaNu
parameters, we follow the Markov Chain Monte Carlo(MCMC) global fit
approach \cite{MCMC} to explore the likelihood space and marginalize
over the nuisance parameters to obtain the constraint on
parameters we are interested in. Our parameter space consists of
\begin{equation}
\vec{P}\equiv (\Omega_bh^2,\Omega_ch^2,H,\tau,A_s,n_s,m_i,\alpha,\beta)~,
\end{equation}
where $\omega_bh^2$ and $\Omega_ch^2$ are the baryon and CDM densities
in units of critical density, $H$ is the hubble parameter, $\tau$ is the
optical depth of Compton scattering to the last scattering surface, $A_s$
and $n_s$ are the amplitude and spectral index of primordial density
fluctuations, and $(m_i,\alpha,\beta)$ are the parameters of MaVaNs.
%%%%%%%%%%%%%%%%%%%%%%%%%%%%%%%%%%%%%%%%%%
\begin{figure}
\begin{minipage}[m]{0.48\linewidth}
    \rotatebox{0}{\includegraphics[width=0.9\textwidth]{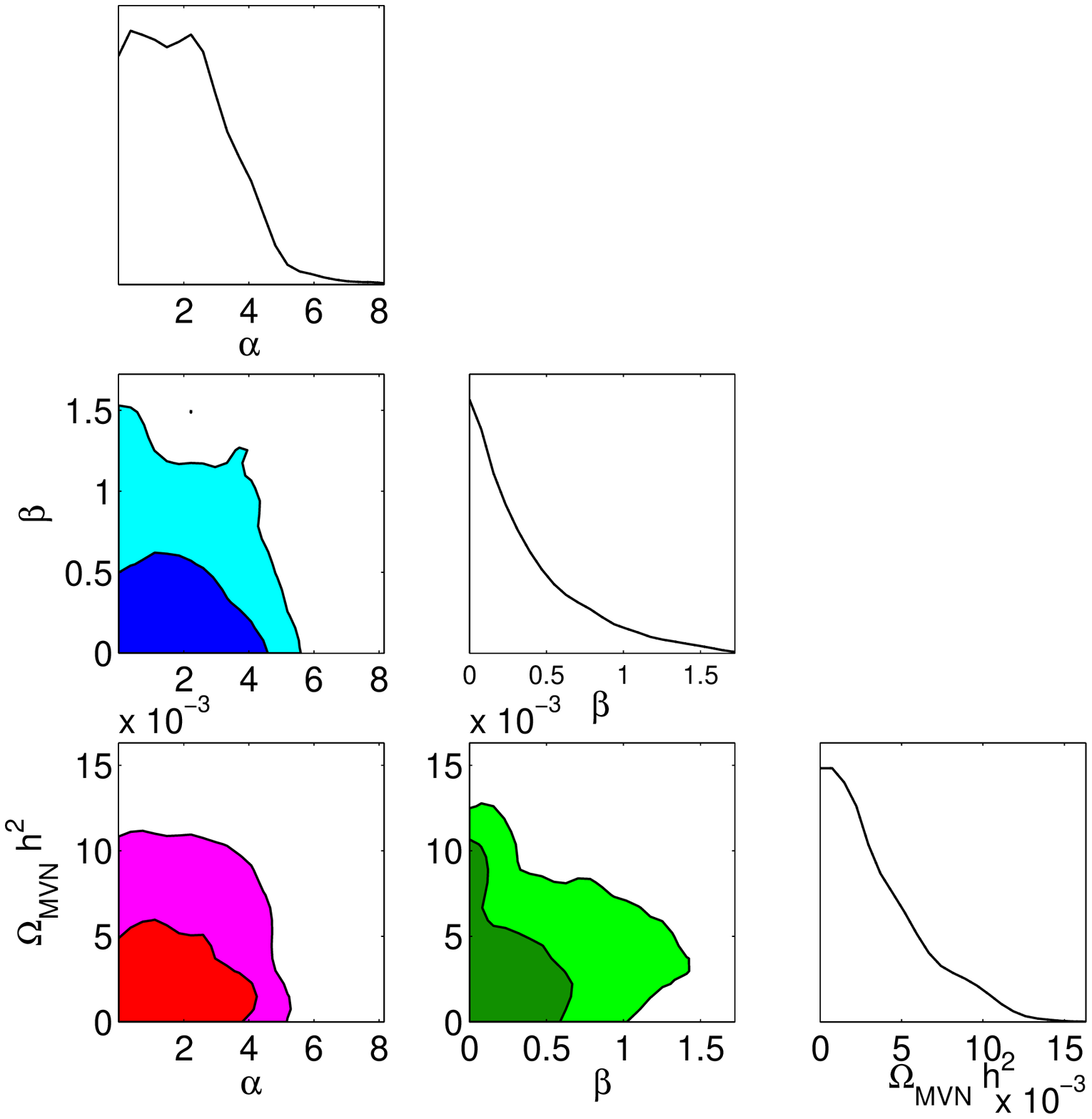}}
\end{minipage}
\begin{minipage}[m]{0.48\linewidth}
    \rotatebox{0}{\includegraphics[width=0.9\textwidth]{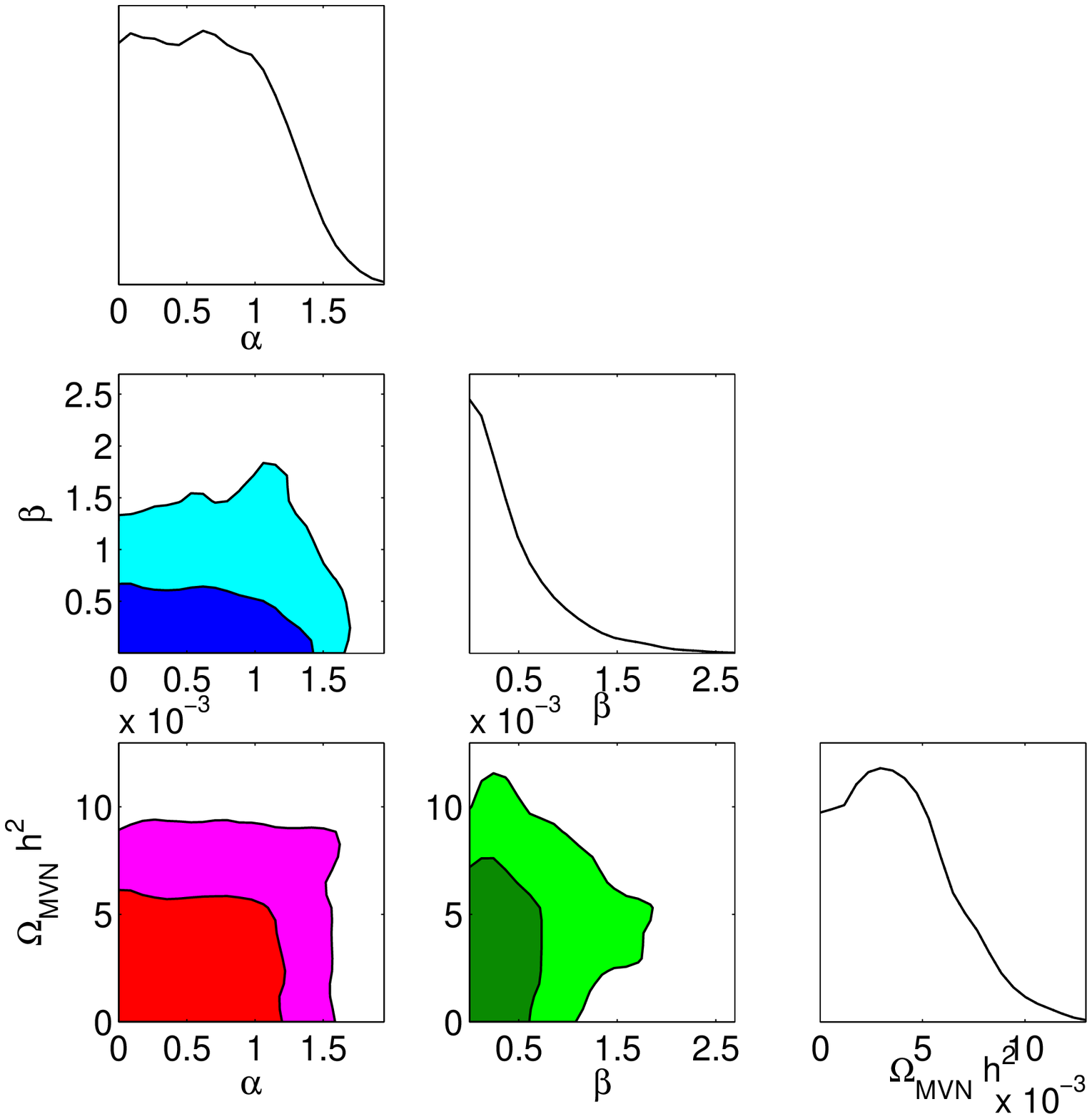}}
\end{minipage}
\caption{(Left panel):Contours of constant relative probabilities in two dimensional
 parameter planes for inverse power law models. Lines correspond to 68\% and 95.4\% confidence limits;
(Right panel):Same as Fig.1-a, but for exponential type models.}
\label{fig:constrained-para}
\end{figure}
%%%%%%%%%%%%%%%%%%%%%%%%%%%%%%%%%%%%%%%%%%
As an example,
allowed parameter's space are shown 
in Figs.(\ref{fig:constrained-para}) for the model I and III.
In these figures we do not observe the strong degeneracy between the
introduced parameters. This is why one can put tight constraints on
MaVaNs parameters from observations. For both models we consider, larger
$\alpha$ leads larger $w$ at present. Therefore large
$\alpha$ is not allowed due to the same reason that larger $w$ is not
allowed from the current observations.
%%%%%%%%%%%%%%%%%%%%%%%%%%%%%%%%%%%%%%%%%%%%%%%%%%%%%%%%%%%%%%%%%%%%%%%%%
\begin{table}[t]
\caption{Global analysis data within $2\sigma$ deviation for different
types of the quintessence potential.}
{\begin{tabular}{c|c|c|c|c} \hline
Quantites & Model I
&Model II & Model III & WMAP-3 data ($\Lambda$CDM)  \\ \hline
$\alpha$ & $< 4.38$ & 0.10 -- 11.82 & $< 1.41$ & ---       \\
$\beta$ & $< 1.12 $ & $< 1.36$ & $< 1.53$ & ---       \\ \hline
$\Omega_B\, h^2[10^2]$ & 2.09--2.36 & 2.09--2.35 & 2.08--2.34 & $2.23\pm 0.07$   \\
$\Omega_{CDM}\, h^2[10^2]$ & 9.87 -- 12.30 & 9.85--12.40 & 9.84--12.33 & $12.8\pm 0.8$ \\
$H_0$ & 58.39 -- 72.10 & 58.55--71.70 & 58.99--71.58 & $72\pm 8$  \\
$Z_{re}$ & 6.13 -- 14.94 & 4.00--14.78 & 6.64--14.78 &   ---     \\
$n_s$ & 0.92 -- 0.99 & 0.92--0.98 & 0.92--0.98 & $0.958\pm 0.016$      \\
$A_s[10^{10}]$ & 18.25 -- 23.41 & 18.20--23.32 & 18.33-23.27 & ---- \\
$\Omega_{Q}[10^2]$ & 57.43 -- 75.60 & 57.59--75.02 & 58.45--75.05 &
$71.6\pm 5.5$      \\
$Age/Gyrs$ & 13.59 -- 14.40 & 13.59--14.35 & 13.61--14.36 &  $13.73\pm 0.16$ \\
$\Omega_{MVN}\,h^2[10^2]$ & $< 0.95$ & $< 0.91$ & $< 0.84$ &  $< 1.97 (95\% C.L.)$ \\
$\tau$ & 0.031--0.143 & 0.028--0.139 & 0.032--0.140 & $0.089 \pm 0.030$  \\ \hline
\end{tabular} \label{tab-1}}
\end{table}
%%%%%%%%%%%%%%%%%%%%%%%%%%%%%%%%%%%%%%%%%%%%%%%%%%%%%%%%%%%%%%%%%%%%%%%%%%
%%%%%%%%%%%%%%%%%%%%%%%%%%%%%%%%%%%%%%%%%%%%%%%%%%%%%%%%%%%%%%%%%%%%%
\begin{figure}
\begin{minipage}[m]{0.48\linewidth}
    \rotatebox{0}{\includegraphics[width=1.0\textwidth]{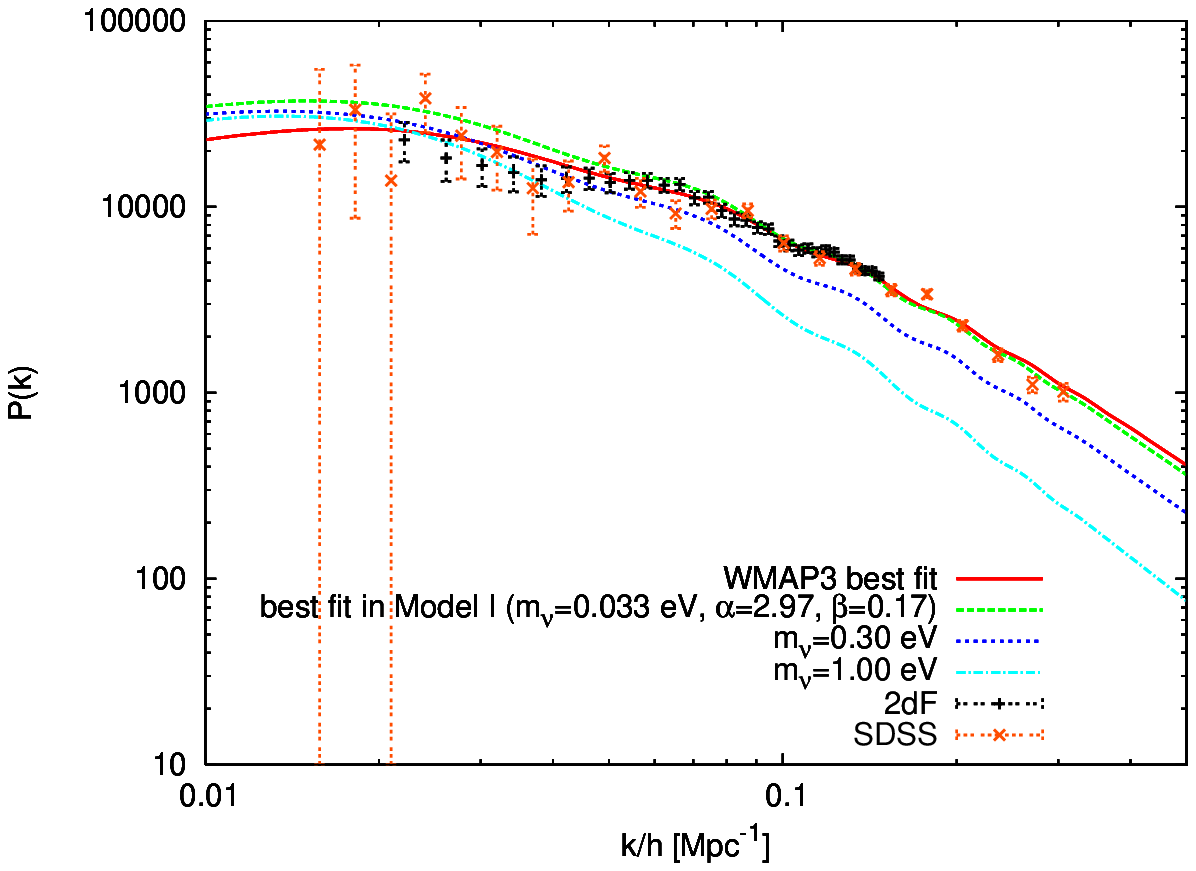}}
\end{minipage}
\begin{minipage}[m]{0.48\linewidth}
    \rotatebox{0}{\includegraphics[width=1.0\textwidth]{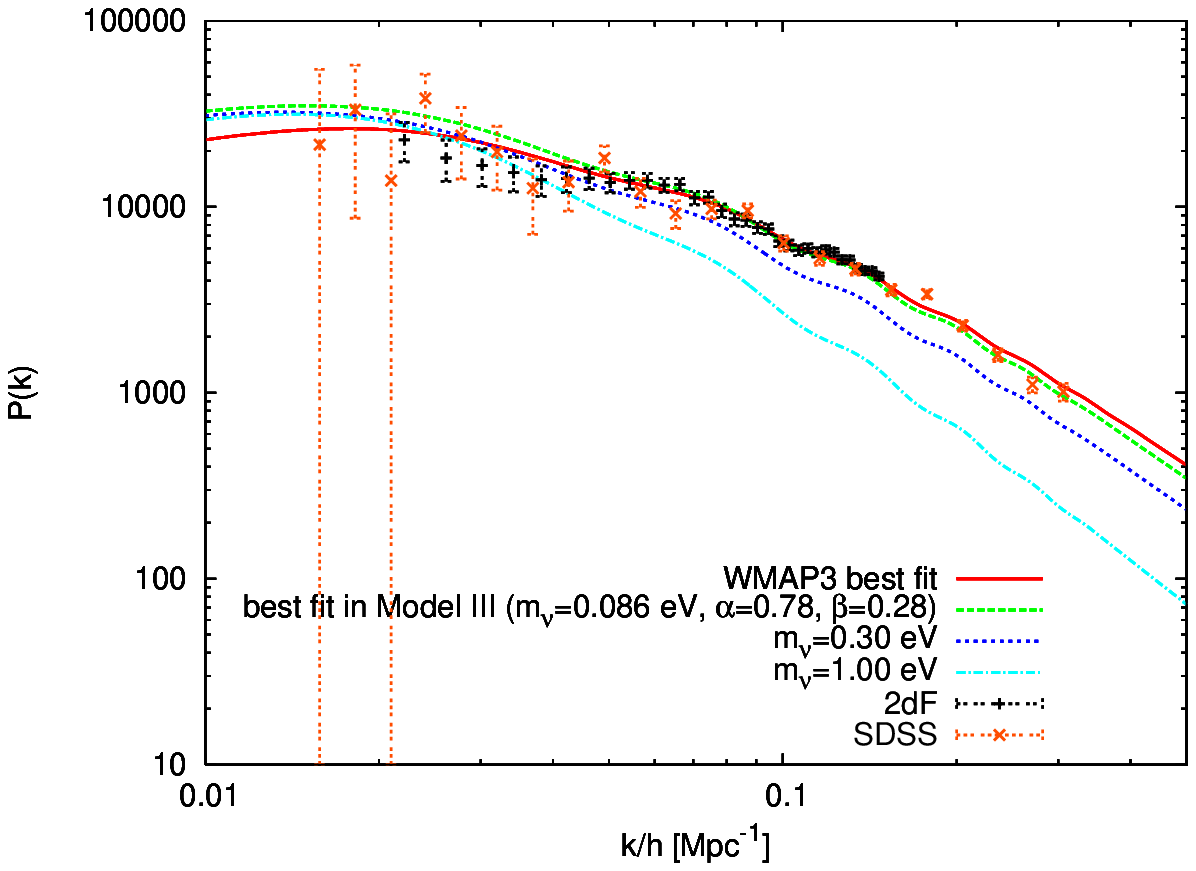}}
\end{minipage}
\caption{Examples of the total mass contributions in
the matter power spectrum in Model I (Left panel) and Model III
(Right panel). For both panels we plot the best fitting lines (green
dashed), lines with larger neutrino masses $M_\nu=0.3$ eV (blue
dotted) and $M_\nu=1.0$ eV (cyan dot-dashed) with the other
parameters fixed to the best fitting values. Note that while lines
with $M_\nu=0.3$ eV can fit to the data well by arranging the other
cosmological parameters, lines with $M_\nu=1.0$ eV can not.
 \label{fig:nu-mass-PS} }
\end{figure}
%%%%%%%%%%%%%%%%%%%%%%%%%%%%%%%%%%%%%%%%%%%%%%%%%%%%%%%%%%%%
We find no observational signature which favors the coupling between
MaVaNs and quintessence scalar field, and obtain the upper limit on the
coupling parameter as shown in table \ref{tab-1}.
\begin{equation}
\beta < 0.46, 0.47, 0.58 \,\,(1 \,\sigma); \,\, 
[1.12,~ 1.36,~ 1.53 \,\,(2 \,\sigma)],
\end{equation}
and the present mass of neutrinos is also limited to
\begin{equation}
\Omega_\nu h^2_{\rm{today}} < 0.0044, ~0.0048, ~0.0048~ \,\,(1\,\sigma); \,\,
 [0.0095,~ 0.0090,~ 0.0084~ \,\, (2\,\sigma)],
\end{equation}
for models I, II and III, respectively.
When we apply the relation
between the total sum of the neutrino masses $M_{\nu}$ and their
contributions to the energy density of the universe:
$\Omega_{\nu}h^2=M_{\nu}/(93.14 eV)$, we obtain the constraint on
the total neutrino mass: $M_{\nu} < 0.45~ eV (68.5 \% C.L.) \,\,[0.87~ eV (95 \% C.L.)$] in the
neutrino probe dark-energy model. The total neutrino mass
contributions in the power spectrum is shown in Fig
\ref{fig:nu-mass-PS}, where we can see the significant deviation
from observation data in the case of  large neutrino masses.
%%%%%%%%%%%%%%%%%%%%%%%%%%%%%%%%%%%%%%%%%%%%%%%%%%%%%%%%%%%%%%%%%%%%%%%%%%
\begin{center}
\begin{figure}[t]
    \rotatebox{0}{\includegraphics[width=0.8\textwidth]{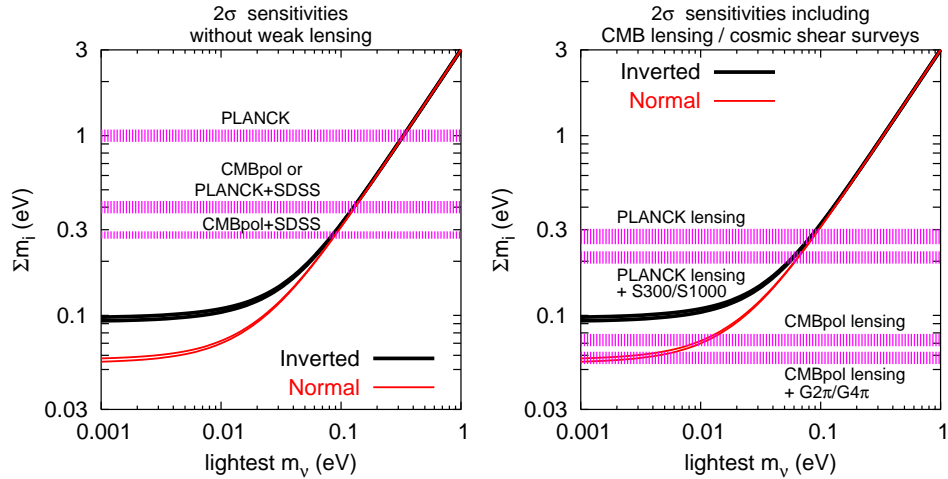}}
\caption{Forecast 2$\sigma$ sensitivities to the
total neutrino mass from future cosmological experiments, 
compared to the values in agreement with
present neutrino oscillation data. 
Left panel: sensitivities
expected for future CMB experiments (without lensing extraction),
alone and combined with the completed SDSS galaxy redshift
survey. Right panel: sensitivities expected for future CMB experiments
including lensing information, alone and combined with future cosmic
shear surveys.  Here CMBpol refers to a hypothetical CMB experiment
roughly corresponding to the Inflation Probe mission. (Fig. from Ref.\cite{julian:2006}).
 \label{fig_future} }
\end{figure}
\end{center}
%%%%%%%%%%%%%%%%%%%%%%%%%%%%%%%%%%%%%%%%%%%%%%%%%%%%%%%%%%%% 
Beyond the scope of our current analysis, there are other possibilities in cosmological probes of neutrino masses:
\begin{itemize}
\item the evolution of cluster abundance with redshit may provide further constraints
on neutrino masses,
\item the Lyman-$\alpha$ forest provides constaints on the matter power spectrum on scale of $k\sim 1\, h\, Mpc^{-1}$, where the effect of massive neutrinos is most viable, 
\item Deep and wide weak lensing survey will make it possible, in the future,
to perform weak lensing tomography of the matter density field.
\end{itemize}
As shown in Fig.\ref{fig_future}, the combination of 
weak lensing tomography and high-precision CMB-polarization experiments may reach
sensitivities down to the lower bound of 0.06 eV on the sum of the neutrino masses
 \cite{{song-knox:2004},{Hannestad:2006as},{julian:2006}}.
In this case, normal hierarchy pattern will be detectable.

\begin{theacknowledgments}
Y.Y.K. would like to thank National Astronomical Observatory in Japan for kind hospitality, and organizers of the 10th International Symposium on Origin of Matter
and Evolution of Galaxies for the invitation at a fruitful and exciting meeting.
Y.Y.K's work is partially supported by Grants-in-Aid for NSC in Taiwan, and Center for High Energy Physics/KNU in Korea. 
K.I's work is supported by Grant-in-Aid for JSPS Fellows.
T.K's work is supported by the Grant-in-Aid for Scientific Research (17540275)
of the Ministry of Education, Culture, Sports, Science and Technology of Japan,
the JSPS Core-to-Core Program, International Research Network for Exotic Femto System (EFES), and the Mitsubishi Foundation.
\end{theacknowledgments}


\begin{thebibliography}{999}
%%%%%%%%%%%%%%%%%%%%%%%%%%%%%%%%%%%%%%%%%%%%%%%%%%%%%%%
\bibitem{solar-data} 
Q.~R. Ahmed et al.[SNO collaboration], Phys. Rev. Lett. {\bf 87}:071301 (2001);
S.~N. Ahmed et al.[SNO collaboration], Phys. Rev. Lett. {\bf 92}:181301 (2004).

\bibitem{super-kam}
S.~Fukuda et al.[Super-Kamiokande collaboration], Phys. Rev. Lett. {\bf 81},
1562 (1998); Phys. Rev. Lett. {\bf 82}, 2644 (1999); Phys. Rev. Lett. {\bf 85},
3999 (2000).

\bibitem{zeldovich}
Ya.~B. Zeldovich and M.~Yu. Khlopov, JETP Lett. (1981) V.34, no. 3, PP. 141-145.

\bibitem{barger:2002}
V.~Barger, S.~L.Glashow, D.~Marfatia, and K.~Whisnant, Phys. Lett. {\bf B532}, 15 (2002).

\bibitem{bilenky:2004}
S.~M. Bilenky, A.~Faessler, and F.~Simkovic, Phys. Rev. {\bf D70}:033003 (2004).

\bibitem{mainz}
J.Bonne et al., Nuch. Phys. B (Proc. Suppl.), 91 (2001) 273.

\bibitem{troitsk}
Ch. Weinheimer, Nucl. Phys. B (Proc. Suppl.), 118 (2003) 279.

\bibitem{heidelberg-moscow}
L. Baudis et al. [Heidelberg-Moscow Collaboration], Phys. Rev. Lett.
{\bf 83}, 41 (1999).

\bibitem{igex}
C.~E. Aalseth et al. [IGEX Collaboration], Phys. Rev. {\bf D 65},
092007 (2002); Phys. Rev. {\bf D 70}, 078302 (2004).

\bibitem{mangano:2005}
G.~Mangano et al., Nucl. Phys. {\bf B729} (2005) 221.

\bibitem{Bilenky:2004} S.~M. Bilenky, A. Faessler, and F. Simkovic,
Phys. Rev. {\bf D 70}:033003 (2004).

\bibitem{Rodin:2003} V.~A. Rodin, A. Faessler, F. Simkovic, and P. Vogel,
Phys. Rev. {\bf C 68}:044302 (2003).

%%%%%%%%%%%%%%%%%%%%%%%%%%%%%%%%%%%%%%%%%%%%%%%%%%%%%%%%

\bibitem{beacom:2004}
J.~F. Beacom, N.~F. Bell and S. Dodelson, Phys. Rev. Lett. {\bf 93}
121302 (2004).

\bibitem{dodelson:2006}
S. Dodelson, A. Melchiorri and A. Slasar, Phys. Rev. Lett. {\bf 97}
04031 (2006).

\bibitem{hannestad:2003}
S. Hannestad, JCAP {\bf 05} 004 (2003).

\bibitem{pierpaoli:2003}
E. Pierpaoli, Mon. Not. Roy. Astron. Soc. {\bf 342} L63 (2003).

\bibitem{hannestad:2005}
S. Hannestad, JCAP {\bf 0502} 011 (2005); arXiv astro-ph/0411475.

\bibitem{slosar:2006}
A Slosar, Phys. Rev. {\bf D73} 123501 (2006).

%\cite{Hu:1997mj}
\bibitem{Hu:1997mj}
  W.~Hu, D.~J.~Eisenstein and M.~Tegmark,
  %``Weighing neutrinos with galaxy surveys,''
  Phys.\ Rev.\ Lett.\  {\bf 80}, 5255 (1998)
  [arXiv:astro-ph/9712057].
  %%CITATION = PRLTA,80,5255;%%


\bibitem{silk:1980}
J.~R. Bond, G. Efstathiou and J. Silk, Phys. Rev. Lett. {\bf 45}
1980 (1980).


%%%%%%%%%%%%%%%%%%%%%%%%%%%%%%%%%%%%%%%%%%%%%%%%%%%%%%%%%%%%
\bibitem{fukujita:2006}
M. Fukugita, K. Ichikawa, M. Kawasaki and O, Lahav, Phys. Rev. {\bf
D 74}, 027302 (2006).

\bibitem{elgaroy:2002}
O. Elgaroy et al., Phys. Rev. Lett{\bf 89} 061310 (2002); O. Elgaroy
and O. Lahav, JCAP 0304 (2003) 004.

\bibitem{sanchez:2005}
A.~G. Sanchez et al., Mon. Not. Roy. Astron. Soc. 366 (2006) 189.

\bibitem{hannestad:2004}
S. Hannestad, JCAP 0305 (2003) 004.

\bibitem{barger:2003}
V. Barger, D. Marfatia, and A. Tregre, Phys. Lett. {\bf B 595} 55
(2004).

\bibitem{spergel:2006}
D.~N. Spergel et al. [WMAP Collaboration], astro-ph/0603449.

\bibitem{goobar:2006}
A. Goodbar, S. Hannestad, E. Mortsell, and H. Tu, J. Cosmol.
Astropart. Phys. 06 (2006) 019.

\bibitem{seljak:2006}
U. Seljak, A. Slosar, and P. McDonald, J. Cosmol Astropart. Phys. 10
(2006) 014.
%%%%%%%%%%%%%%%%%%%%%%%%%%%%%%%%%%%%%%%%%%%%%%%%%%%
\bibitem{ichikawa:2005}
K. Ichikawa, M. Fukugita and M. Kawasaki, Phys. Rev. {\bf D71}
043001 (2005).


%%%%%%%%%%%%%%%%%%%%%%%%%%%%%%%%%%%%%%%%%%%%%%%%%%%%%%%%%%%%%%%%%%%%%%
\bibitem{ichiki-yyk:2007a}
K.~Ichiki and Y.-Y. Keum, arXiv:astro-ph/0705.2134 (to be published in JCAP).

\bibitem{yyk-2007}
Y.-Y.~Keum, Mod. Phys. Lett. {\bf A22}:2131-2142 (2007). 

\bibitem{ichiki-yyk:2008a}
K. ~Ichiki and Y.-Y. Keum, "{\it Neutrino Masses from Cosmological Probes in 
interacting Neutrino Dark-Energy Models}", arXiv:hep-ph/0803.2274.

%%%%%%%%%%%%%%%%%%%%%%%%%%%%%%%%%%%%%%%%%%%%%%%%%%%
\bibitem{mavanu}
D.~B.~Kaplan, A.~E.~Nelson and N.~Weiner, Phys. Rev. Lett. {\bf
93}:091801, (2004); R.~D.~Peccei, Phys. Rev. {\bf D71}:023527
(2005).


%\cite{Fardon:2003eh}
\bibitem{Fardon:2003eh}
R.~Fardon, A.~E.~Nelson and N.~Weiner, JCAP 0410:005, 2004;
%``Dark energy from mass varying neutrinos,''
[arXiv:astro-ph/0309800].
%%CITATION = ASTRO-PH 0309800;%%

%\cite{Bi:2003yr}
\bibitem{Bi:2003yr}
X.~J.~Bi, P.~h.~Gu, X.~l.~Wang and X.~M.~Zhang, Phys. Rev. {\bf
D69}:113007 (2004);
%``Thermal leptogenesis in a model with mass varying neutrinos,''
[arXiv:hep-ph/0311022].
%%CITATION = HEP-PH 0311022;%%

%\cite{Hinshaw:2006ia}
\bibitem{Hinshaw:2006ia}
  G.~Hinshaw {\it et al.}(WMAP collaboration),
  %``Three-year Wilkinson Microwave Anisotropy Probe (WMAP) observations:
  %Temperature analysis,''
  arXiv:astro-ph/0603451.
  %%CITATION = ASTRO-PH 0603451;%%

%\cite{Page:2006hz}
\bibitem{Page:2006hz}
  L.~Page {\it et al.}(WMAP collaboration),
  %``Three year Wilkinson Microwave Anisotropy Probe (WMAP) observations:
  %Polarization analysis,''
  arXiv:astro-ph/0603450.
  %%CITATION = ASTRO-PH 0603450;%%



%\cite{Cole:2005sx}
\bibitem{Cole:2005sx}
  S.~Cole {\it et al.}  [The 2dFGRS Collaboration],
  %``The 2dF Galaxy Redshift Survey: Power-spectrum analysis of the final
  %dataset and cosmological implications,''
  Mon.\ Not.\ Roy.\ Astron.\ Soc.\  {\bf 362}, 505 (2005)
  [arXiv:astro-ph/0501174].
  %%CITATION = ASTRO-PH 0501174;%%


%\cite{McDonald:1999dt}
\bibitem{McDonald:1999dt}
  P.~McDonald, J.~Miralda-Escude, M.~Rauch, W.~L.~W.~Sargent, T.~A.~Barlow, R.~Cen and J.~P.~Ostriker,
  %``The Observed Probability Distribution Function, Power Spectrum, and
  %Correlation Function of the Transmitted Flux in the Lyman-alpha Forest,''
  Astrophys.\ J.\  {\bf 543}, 1 (2000)
  [arXiv:astro-ph/9911196].
  %%CITATION = ASTRO-PH 9911196;%%

%\cite{Croft:2000hs}
\bibitem{Croft:2000hs}
  R.~A.~C.~Croft {\it et al.},
  %``Towards a Precise Measurement of Matter Clustering: Lyman-alpha Forest Data
  %at Redshifts 2-4,''
  Astrophys.\ J.\  {\bf 581}, 20 (2002)
  [arXiv:astro-ph/0012324].
  %%CITATION = ASTRO-PH 0012324;%%


%\cite{Goobar:2006xz}
\bibitem{Goobar:2006xz}
  A.~Goobar, S.~Hannestad, E.~Mortsell and H.~Tu,
  %``A new bound on the neutrino mass from the SDSS baryon acoustic peak,''
  JCAP {\bf 0606}, 019 (2006)
  [arXiv:astro-ph/0602155].
  %%CITATION = ASTRO-PH 0602155;%%

%\cite{McDonald:2004xp}
\bibitem{McDonald:2004xp}
  P.~McDonald, U.~Seljak, R.~Cen, P.~Bode and J.~P.~Ostriker,
  %``Physical effects on the Lyman-alpha forest flux power spectrum: damping
  %wings, ionizing radiation fluctuations, and galactic winds,''
  Mon.\ Not.\ Roy.\ Astron.\ Soc.\  {\bf 360}, 1471 (2005)
  [arXiv:astro-ph/0407378].
  %%CITATION = ASTRO-PH 0407378;%%

\bibitem{MCMC}
A.~Lewis and S.~Bridle,
 Phys. Rev. {\bf D66}, 103511 (2002).

\bibitem{song-knox:2004}
Y.~S. Song and L. ~Knox, Phys. Rev. {bf D 70} (2004) 063510.

\bibitem{Hannestad:2006as}
S.~Hannestad, H.~Tu and Y.~Y.~Y.~Wong, JCAP {\bf 0606}, 025 (2006).

\bibitem{julian:2006}
J. ~Lesgourgues and S. ~Pastor, Phys. Rept. {\bf 429}, 307 (2006).


\end{thebibliography}
\end{document}